\documentclass[a4paper,11pt]{article}
\pdfoutput=1 

\usepackage{jheppub} 

\usepackage[T1]{fontenc} 
\usepackage{appendix}

\usepackage{amssymb,amsfonts}
\usepackage{amsmath}
\usepackage{mathtools}
\usepackage{amsthm}
\usepackage{empheq}
\usepackage[colorinlistoftodos]{todonotes}
\usepackage{array}

\usepackage{graphicx} 
\usepackage{times} 
\usepackage{amssymb}
\usepackage{enumitem}
\usepackage{mathrsfs}
\usepackage{amsmath}
\usepackage{xcolor}

\newcommand{\CL}{{\tt ${\mathcal C}$osmo${\mathcal L}$attice}~}

\newcommand{\be}{\begin{equation}}
\newcommand{\ee}{\end{equation}}
\newcommand{\bea}{\begin{eqnarray}}
\newcommand{\eea}{\end{eqnarray}}

\def\Dslash{\,\,{\raise.15ex\hbox{/}\mkern-13mu D}}
\def\Dbarslash{\,\,{\raise.15ex\hbox{/}\mkern-12mu {\bar D}}}
\def\delslash{\,\,{\raise.15ex\hbox{/}\mkern-10mu \partial}}
\def\delbarslash{\,\,{\raise.15ex\hbox{/}\mkern-9mu {\bar\partial}}}
\def\pslash{\,\,{\raise.15ex\hbox{/}\mkern-11mu p}}
\def\qslash{\,\,{\raise.15ex\hbox{/}\mkern-9mu q}}
\def\kslash{\,\,{\raise.15ex\hbox{/}\mkern-11mu k}}
\def\eslash{\,\,{\raise.15ex\hbox{/}\mkern-9mu \epsilon}}

\newcommand{\slsh}[1]{\,\,{\raise.15ex\hbox{/}\mkern-12mu {#1}}}

\definecolor{myorange}{RGB}{199,146,32}

\def\bs{\ensuremath\boldsymbol}
\newenvironment{aleq}
    {\begin{equation}\begin{aligned}}
    {\end{aligned}\end{equation}\ignorespacesafterend}
\setcitestyle{square}

\title{\boldmath Self-Tracking Solutions for Asymptotic Scalar Fields }

\author[a]{Martin Mosny,}
\author[a]{Joseph P. Conlon,}
\author[b]{Edmund J. Copeland,}

\affiliation[a]{Rudolf Peierls Centre for Theoretical Physics,\\Beecroft Building,\\ Parks Rd, Oxford, OX1 3PU,\\UK}
\affiliation[b]{School of Physics and Astronomy, \\
University of Nottingham, \\
Nottingham, \\
NG7 2RD, UK
}

\emailAdd{martin.mosny@physics.ox.ac.uk}
\emailAdd{joseph.conlon@physics.ox.ac.uk}
\emailAdd{ed.copeland@nottingham.ac.uk}

\preprint{July 2025}

\abstract{ We explore the dynamics of pure scalar fields rolling on an exponential potential in the absence of any additional background fluid and demonstrate the existence of self-tracking solutions in which the self-perturbations of the scalar field act as an effective radiation background. The validity of these solutions is demonstrated through both analytic techniques and numerical simulations using \CL. We discuss applications to string cosmologies with significant trans-Planckian field excursions between inflation and BBN, including the required initial level of scalar perturbations to avoid overshoot.
}

\begin{document}

\maketitle

\section{Introduction}

The history of the universe between the epochs of inflation and nucleosynthesis is largely unconstrained by experiment (e.g. see \cite{Allahverdi:2020bys} for a review). This era offers the opportunity for substantial modification of the Standard Cosmology in which the Hot Big Bang, consisting of a thermal Standard Model radiation bath, follows on immediately after the inflationary epoch.

String cosmologies (for reviews, see \cite{Cicoli:2023opf, Brandenberger:2023ver}) contain many moduli (the scalar fields arising from the compactification geometry). Such moduli can continue to roll in the post-inflationary epoch and the final vacuum may even be located at many Planckian distances in field space from the inflationary locus. 
It is well known that moduli such as the dilaton and volume tend to have exponential potentials that roll towards the asymptotic regions of moduli space \cite{Brustein:1992nk} (as perturbative potentials are power-law in moduli with logarithmic kinetic terms). 
This feature particularly motivates cosmologies in which scalar fields roll substantial distances in the post-inflationary era; a further benefit is that the hierarchies needed for particle physics are easiest to generate in the asymptotic regions of moduli space.

Building on the seminal work of \cite{Ferreira:1997hj, Copeland:1997et}, there exists considerable recent literature \cite{Gouttenoire:2021jhk, Conlon:2022pnx, Revello:2023hro, Shiu:2023fhb, Shiu:2023nph, Seo:2024qzf, Apers:2024ffe, Shiu:2024sbe, Andriot:2024jsh, Conlon:2024uob, Revello:2024gwa, Cicoli:2024yqh, Casas:2024oak, Apers:2022cyl, Apers:2024dtn, Eroncel:2025bcb, Brunelli:2025ems, Andriot:2025cyi, Bedroya:2025ris, Ghoshal:2025tlk, SanchezGonzalez:2025uco} (for older work in a string context see \cite{Barreiro:1998aj, Conlon:2008cj, Cicoli:2015wja}) which explores cosmologies combining a rolling scalar field and a background fluid that may consist of matter, radiation or even more exotic objects such as cosmic strings. In such cases, the final attractor for these cosmologies is often a tracker solution with fixed proportions of the energy density in the background fluid, the kinetic energy of the scalar field and the potential energy of the scalar field.

Although such trackers were originally motivated by the possibility of quintessence in the present universe, in string theory such scenarios face consistency problems due to the twin problems of evolving couplings and unobserved fifth forces. In string theory, such trackers are more appealing as transient epochs in the early pre-BBN universe. They guide the moduli to their final minimum: there, the moduli oscillate and subsequently reheat the universe into the Hot Big Bang.

In this paper, we investigate the possibility that the `background' fluid is itself generated by the self-perturbations of the rolling scalar field. Instead of the case of a scalar field plus a separate background fluid, we explore the more minimal scenario in which the tracker arises entirely from a pure scalar field rolling down an exponential potential. Given the simplicity and ubiquity of such systems, a related goal is to determine the endpoint: does there exist a stable attractor endpoint associated to the combination of the rolling scalar and its own perturbations? (Spoiler: yes).

The paper is organised as follows. Following the introduction (section 1), we give an analytic argument as to why the self-perturbations of a scalar field on an exponential potential can, effectively, act as a radiation background, and can give rise to a tracker solution (section 2). We then describe the results of numerical simulations using \CL that confirm the results of the analytic treatment (section 3). We then discuss physics applications including the possibility that such a self-tracker can address the overshooting problem of string compactifications.

\section{Scalar Self-Tracker: Analytic Treatment}

Rolling scalar fields and tracker solutions have a long history in cosmology \cite{Ratra:1987rm, Copeland:1997et, Ferreira:1997hj} -- for a review on dark energy and the connection to scalar fields, see \cite{Copeland:2006wr}.

In this section we give an analytic description of the tracker formed by a rolling scalar field (on an exponential potential) together with its self-perturbations, which we call the self-tracker.

\subsection{Perturbations in a Scalar Field}

We consider a minimally coupled classical scalar field $\phi(t, \boldsymbol x)$ with potential $V(\phi)$ in a flat FLRW universe with Hubble constant $H$. Although any perturbations that may be present in the field $\phi(t, \boldsymbol x)$ might initially have a quantum origin (such as inflation), our analysis of the dynamics will be fully classical. The equation of motion for the scalar field is given by
\begin{equation}
\label{dtyp}
\ddot \phi - \frac{1}{a^2}\nabla^2 \phi + 3H\dot \phi + \partial_\phi V = 0,
\end{equation}
with its energy and pressure densities given by
\begin{equation}
\begin{split}
\rho_\phi & = \frac{1}{2}\dot \phi^2 + \frac{1}{2a^2}(\nabla \phi)^2 + V(\phi), \\
P_\phi & = \frac{1}{2} \dot \phi^2 - \frac{1}{6a^2} (\nabla \phi)^2 - V(\phi).
\end{split}
\end{equation}
In a universe that is isotropic and homogeneous on large scales, the scalar field can be split into a homogeneous part, referred to as the background, and a inhomogeneous part, referred to as the perturbations, $\phi(t,\boldsymbol x) = \bar \phi(t) + \delta \phi(t,\boldsymbol x)$. These perturbations can themselves be decomposed into Fourier modes
\begin{equation}\label{Fourier}
\delta \phi(t, \boldsymbol x) = \int \frac{d^3 \boldsymbol k}{(2\pi)^3} \delta \phi_{\boldsymbol k}(t)e^{-i\boldsymbol k \cdot \boldsymbol x},
\end{equation}
where reality of the scalar field implies that $\delta \phi_{\boldsymbol k}^* = \delta \phi_{-\boldsymbol k}$. Additionally, any spatial averages over sufficiently large scales result in odd powers of the perturbations vanishing $\langle \delta \phi^{2n+1}\rangle = 0$. If the potential has an exponential form $V = V_0 e^{-\lambda \phi/M_P}$, it can be expanded out as
\begin{equation}\label{eq:pot-expansion}
V(\phi(t,\boldsymbol x)) =V_0 e^{-\lambda \bar \phi(t)/M_P -\lambda \delta \phi/M_P} = \bar V( 1 - \tfrac{\lambda}{M_P} \delta \phi + \tfrac{\lambda^2}{2M_P^2}\delta \phi^2) + \mathcal O(\delta \phi^3),
\end{equation}
provided that the field perturbations are small, $\lambda \delta \phi \ll M_P$. Here $\bar V = V(\bar \phi)$ is defined as the potential of the background field. The density and pressure of the scalar field can in turn be decomposed as
\begin{equation}\label{RhoPressure}
\begin{split}
\rho_\phi & = \bigg(\frac{1}{2}\bar \pi^2 + \bar V\bigg) + \bigg([\bar \pi \delta \pi - \tfrac{\lambda}{M_P}\bar V \delta \phi] + \frac{1}{2}\delta \pi^2 + \frac{1}{2a^2}(\nabla \delta \phi)^2 +\frac{1}{2} \bar m^2\delta \phi^2 + \mathcal O(\delta \phi^3) \bigg),\\
P_\phi & = \bigg(\frac{1}{2} \bar \pi^2 - \bar V\bigg) + \bigg([\bar\pi \delta \pi + \tfrac{\lambda}{M_P} \bar V \delta \phi] + \frac{1}{2}\delta \pi^2 - \frac{1}{6a^2} (\nabla \delta\phi)^2 - \frac{1}{2}\bar m^2 \delta \phi^2 + \mathcal O(\delta \phi^3) \bigg),
\end{split}
\end{equation}
where $\bar m^2 = \lambda^2 \bar V/M_P^2$ is the effective mass of the perturbations and $\pi = \dot \phi$ (which is also decomposed into its homogeneous and inhomogeneous parts). Spatially averaging these gives
\begin{equation}\label{rhoP}
\begin{split}
\langle \rho_\phi \rangle & = \bigg(\frac{1}{2}\bar \pi^2 + \bar V\bigg) + \bigg(\frac{1}{2}\langle\delta \pi^2\rangle + \frac{1}{2a^2}\langle(\nabla \delta \phi)^2\rangle + \frac{1}{2} \bar m^2\langle\delta \phi^2\rangle + \mathcal O(\langle\delta \phi^4\rangle) \bigg), \\
\langle P_\phi\rangle & = \bigg(\frac{1}{2} \bar \pi^2 - \bar V\bigg) + \bigg(\frac{1}{2}\langle\delta \pi^2\rangle - \frac{1}{6a^2} \langle(\nabla \delta\phi)^2\rangle - \frac{1}{2}\bar m^2 \langle\delta \phi^2\rangle + \mathcal O(\langle\delta \phi^4\rangle) \bigg).
\end{split}
\end{equation}
The Hubble equation averaged over these large scales is given by
\begin{equation}\label{eq:Hubble-ave}
H^2 = H^2_{\text{other}} + \frac{1}{3M_P^2}\langle \rho_\phi \rangle,
\end{equation}
where $H^2_{\text{other}}$ represents the spatially averaged contributions to the Hubble equation from any other non-$\phi$ dependent contributions that may be present. The perturbations $\delta \phi(t,\boldsymbol x)$ are assumed to be small enough to avoid any region of space decoupling from the Hubble flow. The scalar field equation of motion to first order can be written as
\begin{equation}\label{SEoM}
\bigg(\ddot{\bar \phi} + 3H \dot{\bar \phi} - \frac{\lambda}{M_P} \bar V \bigg) + \bigg(\delta \ddot \phi + 3H \delta \dot \phi - \frac{1}{a^2}\nabla^2 \delta \phi + \bar m^2 \delta \phi\bigg) + \mathcal O(\delta \phi ^2) = 0.
\end{equation}

Although our focus is on the case where the scalar field is itself responsible for driving the Hubble expansion, we first consider a cosmology where the scalar field is a subleading component to the total energy density and the dominant component has a fixed equation of state $P=w\rho$ such that $a \propto t^c$, with no direct coupling between the dominant energy component and the scalar field. This enables us to see how the scalar perturbations behave within this more simplified cosmology. In this case, the Hubble constant takes the simple form $H=c/t$, where $c = (2/3)/(1+w)$. Working at linear order in the perturbations, we take the spatial average of Eq. \eqref{SEoM} to obtain an equation for the background scalar field
\begin{equation}\label{eq:phibar}
\ddot{\bar \phi} + \frac{3c}{t}\dot{\bar \phi} - \tfrac{\lambda V_0}{M_P} e^{-\lambda \bar \phi/M_P} = 0.
\end{equation}
For $c>1/3$, this equation has an elegant solution where each term scales as $1/t^2$ 
\begin{equation}\label{background_scalar_sol}
\bar \phi(t) = -\frac{M_P}{\lambda} \ln\bigg(\frac{6c-2}{\lambda^2 V_0M_P^{-2}}\bigg) + \frac{2M_P}{\lambda}\ln t,
\end{equation}
during which the effective mass of the perturbations is given by
\begin{equation}
\bar m^2(\bar \phi) = \frac{6c-2}{t^2} = H^2\frac{6c-2}{c^2} \leq \frac{9}{2}H^2,
\end{equation}
with the largest mass occurring during matter domination $c=2/3$. For cosmologies with $1/3 < c \sim \mathcal O(1)$ the mass is of order the Hubble mass $H$.

This solution (\ref{background_scalar_sol}) is a stable fixed point. To see this, consider perturbing around it, $\bar \phi(t) \rightarrow \bar \phi(t)+\delta \bar \phi(t)$. To leading order in the expansion of the exponential, the solution to the linearised version of Eq.~(\ref{eq:phibar}) is
\begin{equation}
\delta \bar \phi(t) = t^{-(3c-1)/2}[c_1\sin(\tfrac{1}{2}\sqrt{(3-c)(9c-3)}\ln t) + c_2 \cos(\tfrac{1}{2}\sqrt{(3-c)(9c-3)}\ln t)],
\end{equation}
for some coefficients $c_1$ and $c_2$. This decays for $c>1/3$, manifestly showing that the late time solution is stable. However, numerically it also appears that the solution of Eq. \eqref{background_scalar_sol} is a late-time attractor for any initial condition. Cosmologies with $c < 1/3$ do not admit an easy physical interpretation.\footnote{For $c<1/3$ the potential falls off too fast to contribute at late times, where the solution instead has the form $\bar \phi(t) = \bar\phi_0 + A t^{1-3c}$.}

The remaining position-dependent part of Eq. \eqref{SEoM} reduces to a linear equation for the perturbations
\begin{equation}\label{eq:pert-eq}
\delta \ddot \phi + 3H \delta \dot \phi - \frac{1}{a^2}\nabla^2 \delta \phi +H^2\frac{6c-2}{c^2}\delta \phi = 0.
\end{equation}
The linearity of Eq.~(\ref{eq:pert-eq}) allows us to directly solve for each Fourier mode. Focusing on one such mode $\delta \phi_{\boldsymbol k}$, its equation of motion is
\begin{equation}\label{perturbation_equation}
\delta \ddot \phi_{\boldsymbol k} + 3H \delta \dot \phi_{\boldsymbol k} + \bigg(\frac{k^2}{a^{2}} +H^2\frac{6c-2}{c^2}\bigg)\delta \phi_{\boldsymbol k} = 0.
\end{equation}
The mass term can be dropped whenever
\begin{equation}\label{MassCondition}
k \gg (aH) \sqrt{\frac{6c-2}{c^2}},
\end{equation} 
a condition which for $1/3 < c \sim \mathcal O(1)$ is roughly equivalent to the perturbations being subhorizon (and all modes will indeed eventually enter the horizon for cosmologies with $c<1$). Plugging in the scale factor $a = a_0(t/t_0)^c$, the sub-horizon modes have an equation of motion given by
\begin{equation}\label{massless_pertubraiton_EoM}
\delta \ddot \phi_{\boldsymbol k} + \frac{3c}{t} \delta \dot \phi_{\boldsymbol k} + \bigg(\frac{kt_0^{c}}{a_0}\bigg)^2 \frac{1}{t^{2c}}\delta \phi_{\boldsymbol k} =0.
\end{equation}
In the case when $c=1/3$ (which corresponds to kination \cite{Spokoiny:1993kt,Joyce:1996cp}, during which the potential is by definition negligible) Eq.~(\ref{massless_pertubraiton_EoM}) is valid for all modes. It can be solved exactly to give (note that $a$ is time-dependent)
\begin{equation}
\delta\phi_{\boldsymbol k}(t) = c_1' t^{-(3c-1)/2}J_\beta\bigg[\frac{ka^{-1}}{1-c} t\bigg] + c_2' t^{-(3c-1)/2}J_{-\beta}\bigg[\frac{ka^{-1}}{1-c} t\bigg],
\end{equation}
where $\beta = (3c-1)/(2-2c)$ and $c_1'$ and $c_2'$ are coefficients fixed by the initial conditions. Using the asymptotic form of the Bessel functions for large real arguments this result can be simplified at either late times or large $k$ to 
\begin{equation}\label{Radiation_perturbation_sol}
\delta\phi_{\boldsymbol k}(t) = c_1 a^{-1}\sin\bigg[\frac{ka^{-1}}{1-c}t\bigg] + c_2 a^{-1} \cos \bigg[\frac{ka^{-1}}{1-c}t\bigg].
\end{equation}

Since $\delta\phi_{\boldsymbol k}(t) \propto a^{-1}$ and $\delta\dot \phi_{\boldsymbol k} \propto a^{-2} + \mathcal O(a^{-3})$, the kinetic, gradient, and potential energy density components scale as
\begin{equation}
\tfrac{1}{2}\delta \pi^2 \propto a^{-4} + \mathcal O(a^{-5}), \ \ \ \ \ \ \tfrac{1}{2a^2}(\nabla\delta \phi)^2 \propto a^{-4}, \ \ \ \ \ \ \ \tfrac{1}{2} \bar m^2 \delta \phi^2 \propto a^{-2-2/c},
\end{equation}
with the potential contribution being negligible at late times whenever $c<1$. This implies that the equation of state is that of radiation $\delta P_\phi = \delta \rho_\phi/3$, a result that holds true in any background cosmology with $a(t) \propto t^c$ when $c \in [1/3,1)$ as long as the scalar field potential is an exponential (for earlier analyses of scalar field perturbations in the case of pure kination, see \cite{Apers:2024ffe, Eroncel:2025bcb}).

The perturbations $\delta \phi$ can be treated as a separate radiation fluid distinct from the spatially-averaged scalar field $\bar \phi$. This is because at leading order, the only direct coupling between the two in the scalar field equation of motion occurs through the mass term $m^2(\bar \phi)\delta \phi = \lambda^2M_P^{-2} V(\bar \phi)\delta \phi$. As long as this term is negligible, as occurs for sub-horizon modes, then the two decouple. Importantly, this is the case irrespective of which component dominates the energy density of the scalar field. In particular, a consistent expansion of the exponential in Eq. (\ref{eq:pot-expansion}) only requires that the inhomogeneities are small relative to the Planck scale $\lambda \delta \phi \ll M_P$, but they need not be small relative to the background scalar field or its time derivative.

While the analysis so far assumed that the scalar field is a subdominant component of the total energy density of the universe, the only way that this assumption came into the picture was to assume that $a \sim t^c$ holds true. The same conclusion holds even when the energy density is dominated by the scalar field itself, and even when $c$ is time-dependent, as long as it does not change substantially on the time scale of the oscillations of the perturbations and one can then use the WKB approximation. As long as the mass $m(\bar \phi)$ is small, the background and the perturbations effectively decouple and the equation of state is then determined by whichever one dominates the energy density of the universe. These dynamics are equivalent to a universe with a radiation fluid and a spatially-independent scalar field. This remains the case throughout the system's evolution as long as the equation of state remains within $c \in [1/3,1)$. 

For sufficiently steep exponential potentials with $\lambda > 2$, this cosmology has a well-known late-time solution given by the radiation tracker for which $c= 1/2$. In this case, such a tracker solution can be made explicit by defining the kinetic and potential energy fraction of the \emph{full} scalar field as \cite{Copeland:1997et}
\begin{equation}
\tilde x = \frac{\dot \phi(t,\boldsymbol x)}{\sqrt 6 HM_P}, \ \ \ \ \ \ \tilde y = \sqrt{\frac{V(\phi)}{3}}\frac{1}{HM_P},
\end{equation}
with the kinetic and potential energy of the \emph{background} field related to these through a spatial averaging
\begin{equation}\label{x_y_definition}
x = \langle \tilde x \rangle = \frac{\bar \pi}{\sqrt 6 HM_P}, \ \ \ \ \ \ y = \sqrt{\frac{V(\bar \phi)}{3}}\frac{1}{HM_P} = \langle \tilde y \rangle[1 + \tfrac{\lambda^2}{8M_P^2}\langle\delta \phi^2\rangle+ \mathcal O(\langle\delta \phi^4\rangle)].
\end{equation}
The energy fraction in the \emph{perturbations} is then given by $z^2 = 1 - x^2 - y^2  = \rho_{\delta\phi}M_P^2/3H^2$, which at late times behaves like radiation with an equation of state $P_{\delta \phi} = \rho_{\delta \phi}/3$. Working with the spatially averaged Hubble parameter of Eq.~(\ref{eq:Hubble-ave}), the familiar autonomous system of differential equations \cite{Copeland:1997et} can be acquired in terms of the number of e-folds $N=\ln a$
\begin{align}
x'(N) & = -3x +\lambda \sqrt{\frac{3}{2}} y^2 + 2x[1+\tfrac{1}{2}x^2-y^2], \\
y'(N) & = -\lambda\sqrt{\frac{3}{2}}xy + 2y[1+\tfrac{1}{2}x^2-y^2], \\
H'(N) & = - 2H[1+\tfrac{1}{2}x^2-y^2], \\
\bar \phi'(N) & = \sqrt 6 M_P x.
\end{align}
where primes represent derivatives with respect to $N$. For sufficiently steep exponential potentials with $\lambda > 2$ this has the standard attractor radiation tracker solution
\begin{equation}
x^2 = \frac{8}{3\lambda^2}, \ \ \ \ \ \ \ \ y^2 = \frac{4}{3\lambda^2}.
\end{equation}

\subsection{Non-linearities and Non-Exponential Potentials}

It is useful to contextualize the case of exponential potentials by considering the case of more general potentials $V(\phi)$ (some of these results have recently been pointed out by \cite{Eroncel:2025qlk}). In that case the scalar field equation of motion can be written as
\begin{equation}
\bigg(\ddot{\bar \phi} + 3H \dot{\bar \phi} + V'(\bar \phi)\bigg) + \bigg( \delta \ddot \phi + 3H \delta \dot \phi - \frac{1}{a^2}\nabla^2 \delta \phi + V''(\bar \phi)\delta \phi\bigg) + \mathcal O\bigg(V'''(\bar \phi)\bigg(\frac{\delta \phi}{\Lambda}\bigg)^2\bigg) = 0,
\end{equation}
where primes denote derivatives with respect to the background field, such as $V'(\bar \phi) \equiv dV(\bar \phi)/d(\bar \phi)$, and where $\delta\phi \sim \Lambda$ is the scale at which the quadratic term in the potential starts to compete with the higher-order terms \cite{Cicoli:2023opf}. If $V''(\bar \phi) \neq 0$, then for sufficiently small perturbations, the higher-order non-linear terms can be ignored. Meanwhile, the Hubble constant depends on both the background scalar field and its perturbations as well as any other sources of energy density that may be present in the universe. If it is sufficiently well described by its spatial average, Eq.~(\ref{eq:Hubble-ave}), then the equation for the perturbations becomes a linear equation. One can then express the perturbations in terms of their Fourier modes, whereby defining a time-dependent frequency $\omega = \sqrt{k^2a^{-2}+ V''(\bar \phi)}$, this equation can in general be solved for sub-horizon modes through the WKB approximation
\begin{equation}
\delta \phi_{\boldsymbol k}(t) = \frac{c_1}{a^{3/2}\sqrt{2\omega}} \sin\bigg(\int^t dt'\omega(t')\bigg) + \frac{c_2}{a^{3/2}\sqrt{2\omega}} \cos\bigg(\int^t dt'\omega(t')\bigg),
\end{equation}
as long as $|\dot \omega/\omega^2|\ll 1$. In particular, whenever $k^2a^{-2} \gg V''(\bar \phi)$, these perturbations behave like radiation.

Crucially, this decomposition does not depend on whether the Hubble constant is dominated by the background $\bar \phi$ or whether it is driven by the perturbations $\delta \phi$. A transition from one to the other does not give rise to non-linearities so long as the quadratic part of the potential continues to dominate over the higher order terms. The only change felt by the perturbations then comes through the Hubble constant. In the case of a quadratic potential, the perturbations behave like a massive scalar field (e.g. see \cite{Eroncel:2025bcb} for a derivation). Meanwhile, when non-linear terms become important, this can give rise to new dynamics such as oscillons (see \cite{Antusch:2017flz, Kasuya:2020szy} for discussions in a string context).

In this work, we have primarily focused on the case of exponential potentials, where the potential factorizes into two exponentials $e^{-\lambda \bar \phi/M_P}e^{-\lambda \delta \phi/M_P}$. The equation for the perturbations can then be expanded out as long as $\lambda \delta \phi\ll M_P$, giving a linear equation, with higher order terms suppressed by factors of $\lambda \delta \phi/M_P$. Additionally, the approximation becomes more accurate with time as long as the amplitude of the perturbations decreases with time. This is true irrespective of whether $\delta \phi$ dominates the energy density of the universe or not. For example, the transition from a kinating epoch, where the energy density is dominated by the kinetic energy in the background field $\dot{\bar \phi}^2/2 \gg V(\phi)$, to one where the universe is dominated by the perturbations, is perfectly linear in terms of the equation of motion for the perturbations $\delta \phi$.

In this work we have neglected metric backreaction from the scalar field, which has been shown to be negligible for adiabatic modes in linear perturbation theory during kination and radiation domination \cite{Eroncel:2025bcb}. The same argument can be extended to the self-tracker to show that the backreaction is negligible in that case as well.\footnote{The only difference from the radiation dominated case of \cite{Eroncel:2025bcb} is the appearance of an additional $2\lambda\bar V \Phi$ term on the right hand side of the scalar field equation of motion (Eq. C.4 in \cite{Eroncel:2025bcb}), which for subhorizon modes scales like $t^{-3}$ and so becomes negligible with time.} It would be interesting to consider non-adiabatic modes, but this is left for future work. If perturbations are large, the metric backreaction can become non-linear and have a significant impact on the evolution of the scalar field modes (see the recent paper \cite{KingsGroup}). This can occur when the energy density in some region is substantially larger than the background $\delta \rho/\bar \rho \gtrsim 1$ for long enough that the region decouples from the Hubble flow \cite{Aurrekoetxea:2024mdy}. In that case a full numerical relativistic analysis is required, with these regions possibly giving rise to black holes. The case of a massless scalar field with an initial Gaussian profile has been studied in numerical relativity \cite{Yoo:2018pda}. This system collapses to a black hole for sufficiently large initial amplitudes, but would otherwise transition into a radiation dominated universe if collapse is avoided, which corresponds in our case to perturbation domination with a negligible potential.

\subsection{Perturbation-Dominated Eras}

It is a familiar feature of radiation tracker solutions that, in the course of their evolution towards the tracker, systems can pass through a transient radiation-dominated phase. The typical course of evolution starts with potential domination, then moves to a kination phase as the field starts rolling down it's potential. The radiation-dominated phase occurs as the radiation catches up and brakes the scalar field; it lasts until the scalar starts rolling again and the tracker is approached.

Since perturbations in scalar fields behave like radiation during a kination phase, it follows that there will be a transient phase where perturbations dominate the Hubble constant. It is not necessarily clear that this era must behave like a radiation domination era. However, on exponential potentials, the Hubble contribution from the background potential is always greater than the contribution from the mass term of the perturbations. This can be shown by using the Fourier decomposition of the perturbations \eqref{Fourier} to write the spatial average of the mass contribution as\footnote{The reason these are single rather than two mode integrals is that the spatial averaging leads to a delta function that eliminates one of the mode integrals.}
\begin{equation}
\frac{1}{2}\bar m^2 \langle \delta \phi^2\rangle = \bar V\frac{\lambda^2}{M_P^2}\int \frac{d^3 \boldsymbol k}{(2\pi)^3} |\delta \phi_{\boldsymbol k}|^2 \leq \bar V\bigg(\frac{\lambda}{M_P}\int \frac{d^3 \boldsymbol k}{(2\pi)^3}|\delta \phi_{\boldsymbol k}|\bigg)^2 \ll \bar V,
\end{equation}
where $|\delta \phi_{\boldsymbol k}| = \sqrt{\delta \phi_{\boldsymbol k}\delta \phi^*_{\boldsymbol k}}$. The first inequality arises from using the integral generalization of the Cauchy–Schwartz inequality, while the second inequality follows from demanding that
\begin{equation}
\frac{\lambda}{M_P}\int \frac{d^3 \boldsymbol k}{(2\pi)^3} |\delta \phi_{\boldsymbol k}| \ll 1,
\end{equation}
which is a marginally stronger requirement than $\lambda \delta \phi/M_P \ll 1$, needed to be able to expand out the exponential to leading order. Since perturbation domination requires the energy density contribution from the perturbations to be much greater than that of the background potential $\bar V$, perturbation domination with small amplitudes cannot behave like a matter dominated era.\footnote{For other potentials, this need not be the case. For example, for quadratic potentials where the background scalar field is at its minimum $\bar V = 0$, one can have a matter dominated era driven by the energy density coming from the perturbations around the minimum.}

Any perturbation dominated era must therefore be driven by the kinetic and gradient terms of the perturbations. It follows from Eq. \eqref{rhoP} that for the energy density to be dominated by the perturbations, it must be the case that
\begin{equation}
\frac{1}{2}\int \frac{d^3 \boldsymbol k}{(2\pi)^3} \bigg(|\delta \dot \phi_{\boldsymbol k}|^2  +  \frac{k^2}{a^2} |\delta \phi_{\boldsymbol k}|^2\bigg) \gg \bar V \gg  \frac{1}{2}\int \frac{d^3 \boldsymbol k}{(2\pi)^3}\bar m^2 |\delta \phi_{\boldsymbol k}|^2.
\end{equation}
A necessary, but not sufficient, condition for the kinetic terms to dominate requires $|\delta \dot \phi_{\boldsymbol k}| \gg \bar m|\delta \phi_{\boldsymbol k}|$, while for gradient terms to dominate requires $ka^{-1}|\delta \phi_{\boldsymbol k}| \gg \bar m |\delta \phi_{\boldsymbol k}|$, at least for the most relevant Fourier modes.

We can now check that in either of these cases, the mass term is also a subdominant term in the equation of motion, given by
\begin{equation}
\delta \ddot \phi_{\boldsymbol k} + 3H \delta \dot \phi_{\boldsymbol k} + \big( \tfrac{k^2}{a^2} + \bar m^2 \big)\delta \phi_{\boldsymbol k} = 0.
\end{equation}
For large gradients, $k/a\gg \bar m$ directly implies that the gradient term must dominate over the mass. Meanwhile, we can estimate the magnitude of the Hubble term to be
\begin{equation}
3H |\delta \dot \phi_{\boldsymbol k}| \gg 3H\bar m|\delta \phi_{\boldsymbol k}| \gg \frac{1}{\lambda}\bar m^2 |\delta \phi_{\boldsymbol k}| \sim \bar m^2 |\delta \phi_{\boldsymbol k}|,
\end{equation}
where for the second inequality we have used that during perturbation domination $H^2 \gg \bar V/M_P^2 = \bar m^2/\lambda^2$. Therefore, for $\lambda \sim \mathcal O(1)$, the mass term is much smaller than the Hubble term. We have thus shown that when the mass contribution to the energy density is much smaller than the gradient/kinetic terms, then the gradient/Hubble terms dominate over the mass term in the equation of motion, respectively. The equation therefore corresponds to that of a massless scalar field, implying that during perturbation domination with subhorizon modes, the universe behaves like a radiation era.

This is not to say that perturbations on an exponential potential can never behave like a massive scalar field. For example, if the background is frozen at some value of the potential $\bar V \sim$ constant, then the equation for the perturbations is
\begin{equation}
\delta \ddot \phi_{\boldsymbol k} + 3H\delta \dot \phi_{\boldsymbol k} + \bigg(\frac{k^2}{a^2} + \frac{\lambda^2 \bar V}{M_P^2}\bigg) \delta \phi_{\boldsymbol k} = 0,
\end{equation}
which for $k/a < \lambda \sqrt{\bar V}/M_P$ behaves like a massive scalar field. In a cosmology with $H = c/t$, this can be solved to give
\begin{equation}
\delta \phi_{\boldsymbol k} = c_1t^{-\beta}J_{\beta}(mt) + t^{-\beta}Y_{\beta}(mt) \approx c_1'a^{-3/2}\sin(mt) + c_2' a^{-3/2}\cos(mt), 
\end{equation}
where $\beta = (3c-1)/2$, and in the second step we used the asymptotic form of the Bessel functions for large $mt$. This solution behaves like a matter field with mass $m = \lambda \sqrt{\bar V}/M_P$. However, so long as the background scalar field can start rolling down the exponential, then such matter behavior would be short-lived.

\subsection{The Final Minimum and Matter Domination}

In any physically realistic scenario, the moduli must avoid overshooting and must instead end up settling down in the final minimum of the potential. As the field approaches the minimum, the potential must transition from an exponential potential to a quadratic one, approximated near the minimum as $V(\phi) = m^2 \left( \phi - \phi_f \right)^2$, where $\phi_f$ is the final value of the scalar field at the minimum. In that case the background field satisfies
\begin{equation}
\ddot{\bar \phi} + 3H \dot{\bar \phi} + m^2 \bar \phi = 0,
\end{equation}
while the perturbations satisfy
\begin{equation}
\delta \ddot \phi_{\boldsymbol k} +3H\delta \dot \phi_{\boldsymbol k}  + \bigg(\frac{k^2}{a^2} + m^2\bigg)\delta \phi_{\boldsymbol k} = 0.
\end{equation}
As the scalar field oscillates around the minimum, the equation of state for the universe is that of a matter dominated era $a\propto t^{2/3}$.\footnote{As it oscillates, the background scalar field and the Hubble constant take the form
\begin{equation}
\bar \phi(t) = \frac{\bar \phi_0}{t}\sin(mt)+\cdots, \ \ \ \ \ H = \frac{2}{3t}+\frac{A\sin^2(\theta_0+mt)}{t^2}+\cdots,
\end{equation}
for some coefficient $A$ and phase $\theta_0$.}

If the transition from the exponential potential to the quadratic potential is approximately instantaneous, then we can estimate the change in amplitude of the perturbations as Fourier modes transition from behaving like radiation to behaving like matter. Consider a Fourier mode with wavenumber $k$ and amplitude $|\delta \phi_{\boldsymbol k,i}|$ at the end of the exponential era when the scale factor is $a_i$. If $k/a_i<m$ then this mode will immediately transition into a massive mode with an amplitude subsequently scaling as $|\delta \phi_{\boldsymbol k}| = |\delta \phi_{\boldsymbol k,i}|a^{-3/2}$. If instead $k/a_i>m$, then this mode behaves like a massless mode until it becomes non-relativistic at $a_c = k/m$. Therefore, these amplitudes scale as
\begin{equation}
|\delta \phi_{\boldsymbol k}| = \begin{cases}
|\delta \phi_{\boldsymbol k,i}|a^{-1},  \ \ \ \ \ \ \ & a<a_c,\\
|\delta\phi_{\boldsymbol k,i}|\sqrt{a_c} a^{-3/2}, \ \ \ \ \ \ \ &a>a_c.
\end{cases} 
\end{equation}

In the case of a cosmology that reached a self-tracker solution before the minimum, the transition into a quadratic potential will dilute the relative contribution of the perturbations to the total energy density due to some of them diluting like radiation in a matter dominated era. They will keep diluting until the modes become non-relativistic, after which the fractional energy density in the perturbations is fixed relative to the energy density in the background $\bar \phi(t)$, since they both scale as matter.

\section{Scalar Self-Tracker: Numerical Treatment}

\subsection{Simulation Description}

In this section, we analyze these systems numerically. We do so using the publicly available code \CL \cite{Figueroa:2020rrl, Figueroa:2021yhd} to simulate a scalar field with an exponential potential in a flat FLRW universe (to be clear, \CL does a full numerical simulation of the scalar field but does \emph{not} simulate the metric; it does not do numerical GR). A full numerical GR simulation of a kinating scalar field will appear shortly in \cite{KingsGroup}.

We use a $32^3$ sized lattice, which is sufficient for us to keep our errors under $0.1\%$ at all times.
In these simulations, the Hubble constant is given in terms of the average energy density in the simulation box. For the simulation, \CL transforms the physical spacetime variables $\{\phi, dt, dx\}$ into dimensionless program variables $\{\tilde \phi, d\tilde \eta, d\tilde x\}$ through
\begin{equation} \label{program_units}
\phi = \tilde \phi/f_*, \ \ \ \ \ \ \ d\eta = \omega_* a^{-\alpha} dt, \ \ \ \ \ \ d\tilde x = \omega_* dx,
\end{equation} 
where both $\omega_*$ and $f_*$ have dimensions of energy and can be chosen arbitrarily to make variables of order one in program units. The choice of $\alpha$ depends on the particular physical situation being simulated, and should be chosen at the start such that the program time tracks with the smallest relevant temporal time scale to avoid numerical instability at late times. Other quantities such as the kinetic, gradient, and potential energies and the Hubble constant are also appropriately rescaled to make them dimensionless.

For simplicity, we restrict ourselves to simulating scalar fields with a single standing wave in comoving space. When the oscillation frequency is much faster than the rate of change of the equation of state of the universe, we expect the subhorizon perturbations to behave like a massless scalar field, with a temporal component given by Eq.~\eqref{Radiation_perturbation_sol}. Writing the Hubble constant as $H=c/t$, the scalar field has the form 
\begin{equation}\label{Initial_Scalar_Field}
\phi(t, \boldsymbol x) = \bar \phi(t) + A(t) \cos\bigg(\frac{ka^{-1}}{1-c} t\bigg)\sum_{j=1}^3\cos(k x_j),
\end{equation}
where the wavenumber is given by
\begin{equation}
k = \frac{2\pi N_{\tilde \lambda}}{L} = k_{IR}N_{\tilde \lambda}.
\end{equation}
Here $L=2\pi/k_{IR}$ is the physical size of the box being simulated, which is converted into the corresponding infrared cutoff $k_{IR}$. Meanwhile, $N_{\tilde \lambda}$ is the number of standing waves present in the box. The smallest relevant physical timescale is the oscillation period of the scalar field perturbations, which at time $t$ is given by\footnote{This is found by expanding Eq.~\eqref{Initial_Scalar_Field} around $t \rightarrow t + \Delta t$ for $\Delta t \ll t$, noting that $a^{-1}t \propto t^{1-c}$.} 
\begin{equation}
\Delta t \approx \frac{2\pi a(t)}{k}.
\end{equation}
To resolve each oscillation with the same number of program time steps we need $d\eta$ to scale like $\Delta t$, which requires setting $\alpha = 1$. Since e-folds scale with time as $dN\propto a^{1-1/c} d\eta$, this implies that each successive e-fold takes longer to simulate than the previous one in cosmologies with $c<1$. To ensure that perturbations behave like radiation in non-kination epochs, we want them to be sub-horizon. We therefore set the initial ratio of the comoving wavelength $\tilde \lambda$ to the comoving Hubble length $r_H = (aH)^{-1}$ to be comparable to unity $(\tilde \lambda r_H^{-1})_i\lesssim1$. In cosmologies with $c<1$, with time they become more sub-horizon since the Hubble horizon grows in comoving space
\begin{equation}
\tilde \lambda r_H^{-1} = (\tilde \lambda r_H^{-1})_i a^{1-1/c}.
\end{equation}

To find suitable initial conditions at time $t=t_i$, where we set $a(t_i)=1$, we pick $\phi(t=t_i,\boldsymbol x)$ and $\dot \phi(t=t_i,\boldsymbol x)$ corresponding to a desired split in the energy fractions between the background kinetic energy, the background potential energy, and the energy in the perturbations. One way to do this is to place all the initial energy of the perturbations into the gradient term. After performing a spatial average, the gradient energy is given by
\begin{equation}
E_{g,i} = \frac{1}{2} \sum_{j=1}^3 \langle (\partial_j \phi)^2\rangle = \frac{3}{4}A_i^2 N_{\tilde \lambda}^2k_{IR}^2,
\end{equation}
where $A(t_i) = A_i$. This can be expressed in terms of the perturbation energy fraction $z$ using the initial Hubble constant $H_i$ as
\begin{equation}
z^2_i = \frac{E_{g,i}}{3 H_i^2M_P^2}.
\end{equation}
The initial background scalar field kinetic energy fraction $x$ and the potential energy fraction $y$ are given by
\begin{equation}
x^2_i = \frac{\bar \pi_i^2}{6H_i^2M_P^2}, \ \ \ \ \ \ \ y^2_i = \frac{V_i e^{-\lambda \bar \phi(t_i)/M_P}}{3 H_i^2M_P^2}.
\end{equation}
Desired initial conditions can then be acquired by picking appropriate $A_i$, $\bar \pi_i$, and $\bar \phi(t_i)$ for some initial Hubble constant $H_i$.

It is important to note that the simulation itself uses the full scalar field $\phi(t,\boldsymbol x)$ and does not perform the split into background and perturbation components at any point. At any given time slice, the simulation outputs spatial averages of the full quantities. We perform the split into background and perturbation components after the simulation, by taking the position independent part of the scalar field as the spatial average of the full scalar field $\bar \phi(t) = \langle \phi(t,\boldsymbol x)\rangle$, while the square of the perturbations is given by the variance of the field
\begin{equation}
\langle \delta \phi(t,\boldsymbol x)^2\rangle = \langle \phi(t,\boldsymbol x)^2\rangle - \langle \phi(t,\boldsymbol x)\rangle^2.
\end{equation}
The same decomposition can be performed for the time derivative of the scalar field. The simulation similarly outputs the full, but spatially averaged, kinetic energy $E_k$, gradient energy $E_g$, and potential energy $E_p$ at each time slice. From these we extract the energy fractions by defining the kinetic energy of the background scalar field directly from its spatially-averaged velocity
\begin{equation}
x^2 = \frac{\bar \pi^2}{6 H^2M_P^2}.
\end{equation}
The energy in the perturbations can then be defined as the sum of the excess total kinetic energy, gradient energy, and potential energy compared to those of the spatially averaged values.
\begin{equation}
z^2 = \bigg(\frac{E_k}{3H^2M_P^2} - x^2\bigg) + \frac{E_g}{3H^2M_P^2} + \frac{\lambda^2 \bar V \langle \delta \phi^2\rangle}{6H^2M_P^4}.
\end{equation}
The potential energy of the background field is given by
\begin{equation}
y^2 = \frac{E_p}{3H^2M_P^2} - \frac{\lambda^2 \bar V \langle\delta \phi^2\rangle}{6H^2M_P^4}.
\end{equation}
Since $E_p$ is defined from the position average of the potential with the full scalar field $\langle V(\phi)\rangle$ rather than the potential of the averaged scalar field $V(\langle \phi\rangle)$, this definition differs from the one in Eq. \eqref{x_y_definition} by $O(\delta \phi^4)$ corrections.

The numerical error in the simulation is estimated by looking at the relative deviation in the spatially averaged Hubble equation \cite{Figueroa:2021yhd}
\begin{equation}
r_E = \frac{\langle LHS- RHS\rangle}{\langle LHS + RHS \rangle}.
\end{equation}
where we defined
\begin{equation}
LHS = \bigg(\frac{da}{d\eta}\bigg)^2, \ \ \ \ \ \ \ RHS = \frac{ a^{2\alpha + 2}}{3M_P^2\omega_*^2}(E_k+E_g+E_p).
\end{equation}

\subsection{Simulation Results}

We run two illustrative simulations to show that the cosmological evolution of a scalar field on an exponential potential, with a perturbation of a single Fourier mode, behaves like a homogeneous scalar field together with a background radiation fluid. The parameter choices and initial conditions in these simulations are given in Table~\ref{Table}. We start both simulations with modes close to the Hubble scale, which expedites the simulation since then we have to simulate fewer oscillations per e-fold. We also ensure that in both simulations the error stays below $0.1\%$ at all times and only run the simulations for eight e-folds.

\begin{table}[b]
\centering
\label{tab:parameters}
\begin{tabular}{||c | c c | c c |c c c ||}
 \hline
 Figure & $\lambda$ & $N_{\tilde \lambda}$ & $H_i$ (GeV) & $(\tilde \lambda r_H^{-1})_i$ & $x^2_i$ & $y^2_i$ & $z^2_i$ \\ [0.5ex] 
 \hline
 $1$ & $3$ & $2$ & $4.5 \times 10^8$ & $1.4$ & $0.495$ & $0.498$ & $0.007$ \\ 
 \hline
 $2$ & $3$ & $2$ & $5.5 \times 10^7$ & $0.17$ & $0.23$ & $0.16$ & $0.61$ \\
 \hline
\end{tabular}
\caption{Displays the values of parameters and initial energy fractions used in the simulations.}
\label{Table}
\end{table}

\begin{figure}[htp]
    \centering
    \includegraphics[width=12cm]{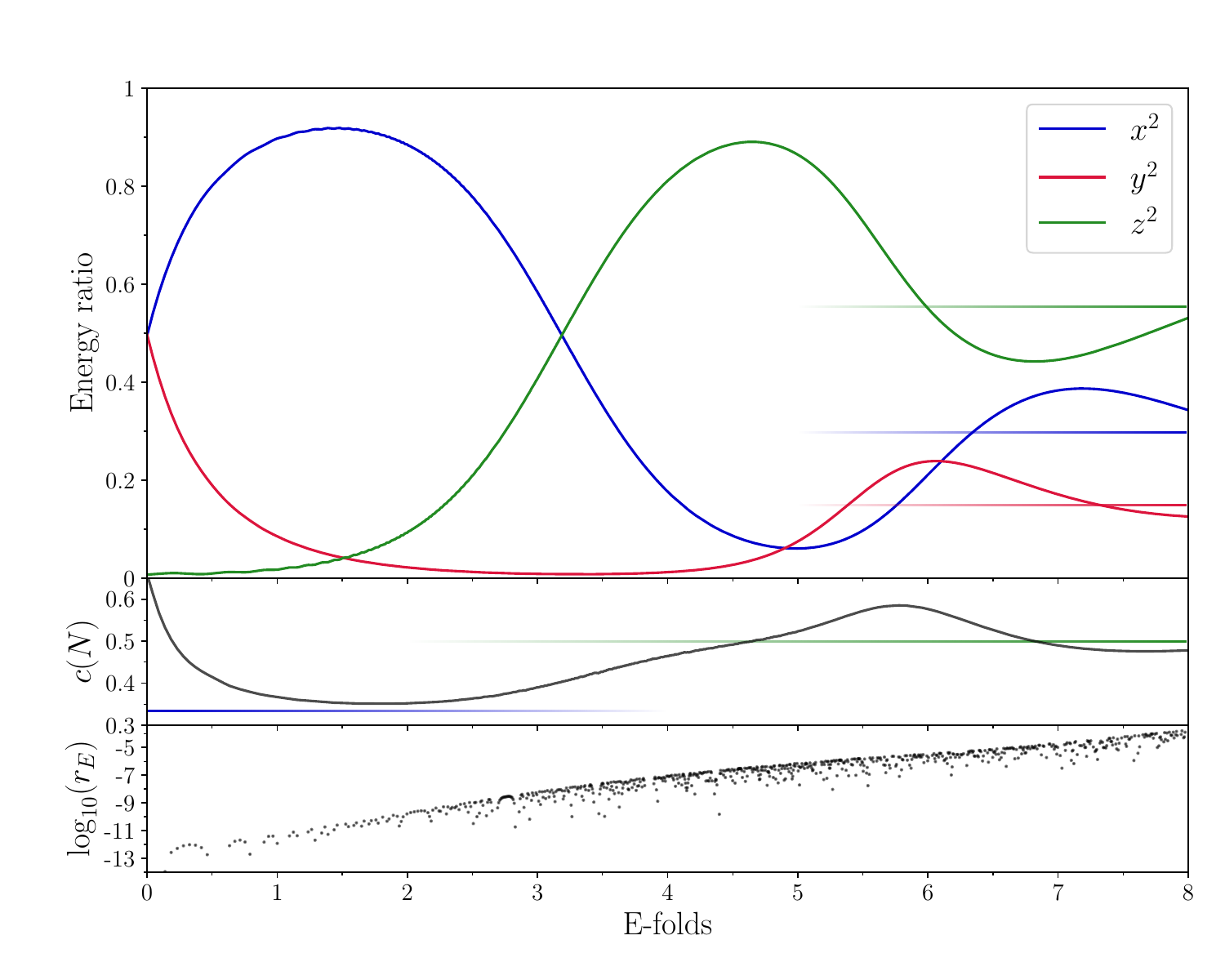}
    \caption{The top panel plots the energy fraction in the background scalar field kinetic energy (blue) and potential energy (red), and the energy in the scalar field perturbations (green) over eight e-folds. The horizontal lines on the right indicate the radiation tracker values. The middle panel displays the local value of $c$, indicating the equation of state of the universe, with a blue horizontal line for $c=1/3$ and a green one for $c=1/2$. The bottom panel plots the normalized error in the Hubble equation $r_E$, showing that this stays below $10^{-3}$.}
    \label{fig:Evolution}
\end{figure}

\begin{figure}[htp]
    \centering
    \includegraphics[width=12cm]{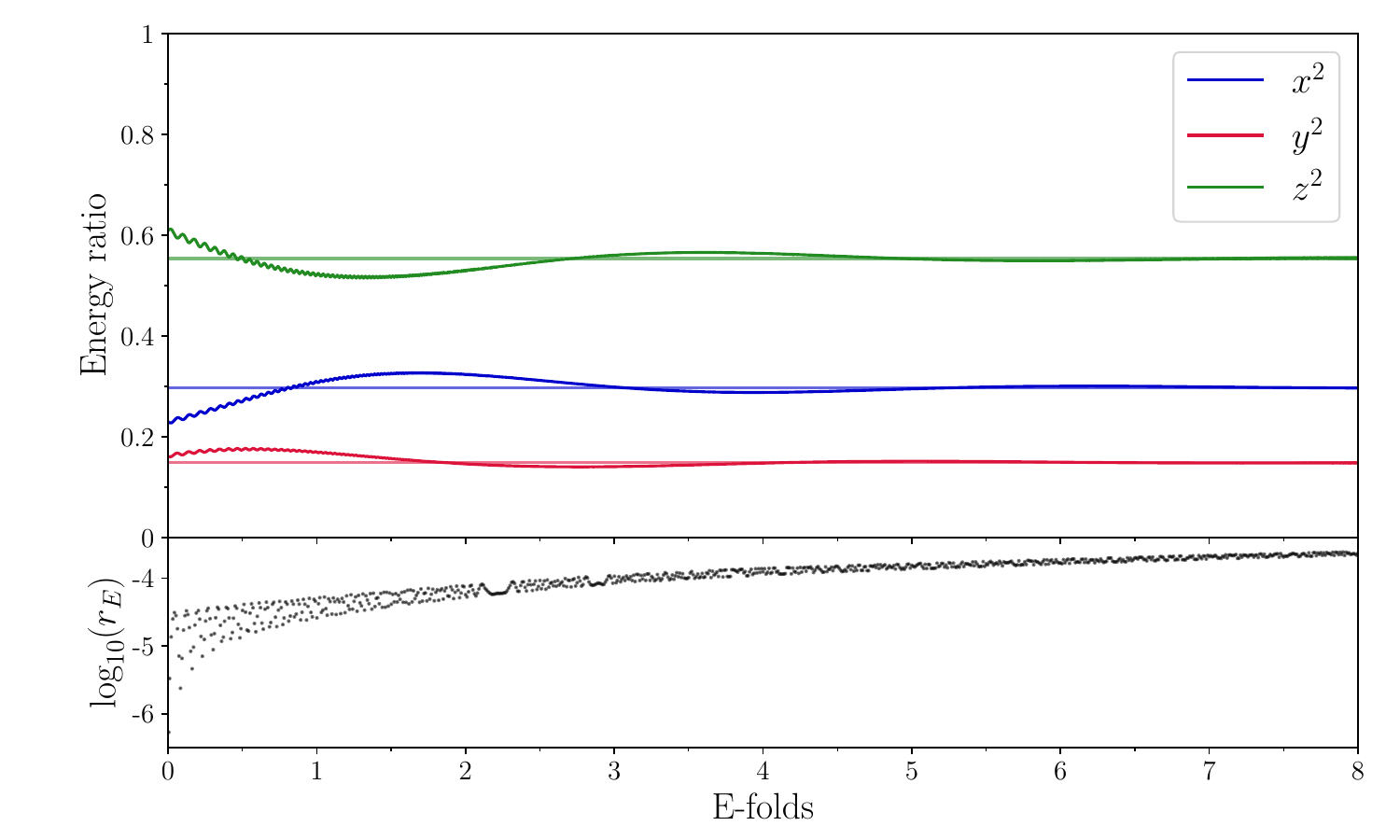}
    \caption{The top panel displays the settling down of the dynamical system into the self-tracker, while the bottom panels displays normalized error in the Hubble equation $r_E$ for this evolution. The horizontal lines indicate the radiation tracker values.}
    \label{fig:RingDown}
\end{figure}

The results of the first simulation are shown in Figure~\ref{fig:Evolution}. This simulation starts out with $x^2 \approx  y^2 \gg z^2$. The cosmology follows the expected evolution, with the scalar field initially rolling down the exponential and then entering a kination era where the kinetic energy dominates. This transitions into a radiation domination era once the perturbations in the scalar field catch up and dominate. Once the potential energy becomes comparable to the other two components, the system enters the self-tracker regime. The middle panel show the local value of $c$, which can be calculated through
\begin{equation}
c = -\frac{H^2}{\dot H}.
\end{equation}
This approaches $1/3$ during kination domination, $1/2$ during perturbation domination, and then after a temporary departure from this value during the transition into the self-tracker, it starts to oscillate and settled down to $1/2$.

The results of the second simulation are presented in Figure~\ref{fig:RingDown}. This shows a simulation starting close to the energy fractions of a radiation tracker and illustrates how the cosmology settles down into exactly those values. There are two types of oscillations of note in Figure~\ref{fig:RingDown}. First, there are oscillations in the energy fractions at early times, which are a consequence of the choice to start out with modes close to the Hubble horizon. In particular, from Eq. \eqref{Radiation_perturbation_sol} it follows that the velocity in the perturbations for $c_{2,\boldsymbol k} = 0$ and $c_{1,\bs k}=A$ is given by
\begin{equation}
\delta \pi_{\boldsymbol k} = \frac{Ak}{a^2}\cos\bigg(\frac{ka}{1-c}\bigg) - \frac{AH}{a}\sin\bigg(\frac{k a}{1-c}\bigg).
\end{equation}
For horizon size modes $k \sim aH$ the two terms are comparable, which yields an oscillatory cross-term in the kinetic energy contribution to $H^2$. As the modes become more sub-horizon, these oscillations get suppressed. These are also seen to a lesser extent in Figure~\ref{fig:Evolution}, with their impact being suppressed due to the lower relative contribution of the perturbations to the total energy density of the simulation at early times. 

The second type of oscillations present are the oscillations around the radiation tracker as the solution settles down into the tracker values. Following the analytic treatment of \cite{Apers:2024ffe}, these are expected to have a frequency of 
\begin{equation}
\tilde k^2 = \frac{15}{4} - \frac{16}{\lambda^2}.
\end{equation}
In our case this corresponds to an oscillation roughly every $\Delta N = 4.47$, which is reproduced by the simulation.

\section{Discussion and Conclusion}

\subsection{Duration of Kination}

If inflation is followed by a period of kination, then this period can only last as long as it takes for the perturbations to catch up with the background kinetic energy, unless there is another source of matter present that can start to dominate first. After a period of perturbation domination, the universe would then enter the cosmological self-tracker which would guide the modulus to the final minimum. Whether this scenario is indeed feasible depends on whether there is sufficient time after inflation for perturbations in the scalar field to catch up with the background before the scalar field settles down into its minimum (otherwise, the overshoot problem rears its head \cite{Brustein:1992nk}). During this era, the energy fraction in the perturbations is given by
\begin{equation}
\Omega_\delta(k) = \langle \delta \rho_\delta / \bar \rho\rangle,
\end{equation}
with kination terminating once $\Omega_{\delta}(k) \approx 1$. Since the perturbations behave like radiation, their energy density \eqref{RhoPressure} can be written as\footnote{This follows because for radiation $\delta \pi^2/2 = (\nabla \delta \phi)^2/2a^2$.}
\begin{equation}
\delta \rho_{\delta}/\bar \rho = (\rho - \bar \rho)/\bar \rho = \bar \pi \delta \pi/\bar \rho + \delta \pi^2/\bar \rho = \delta + \delta^2/2,
\end{equation}
where $\bar \rho = \bar \pi^2/2$ is the kination energy density. Meanwhile, $\delta = 2\delta \pi/\bar \pi$ is the leading contribution, leading in the perturbation, which vanishes upon a spatial averaging, leaving behind $\Omega_{\delta}(k) = \tfrac{1}{2}\langle \delta^2\rangle$. This can be expressed in terms of the dimensionless power spectrum sourced by modes entering the horizon, from the time when kination begins ($k_{\text{kin}}$) until a later time ($k = aH$)\footnote{The power spectrum is that for the linear perturbation $\mathcal P_{\delta}(k) = \int d^3 \boldsymbol x \langle\delta(0)\delta(\boldsymbol x)\rangle e^{i\boldsymbol k \cdot \boldsymbol x}$.}
\begin{equation}
\Omega_\delta(k) = \frac{1}{2}\int_k^{k_{kin}}d \ln \tilde k \ \mathcal P_{\mathcal \delta}(\tilde k).
\end{equation}

If these perturbations are sourced solely by inflation, then their power spectrum can be directly related to the curvature power spectrum from inflation \cite{Eroncel:2025bcb}
\begin{equation}
\mathcal P_\delta(k) = \frac{8}{\pi}\frac{k}{aH}\mathcal P_{\mathcal R}(k).
\end{equation}
Assuming that the primordial curvature power spectrum is sufficiently well approximated by the power law
\begin{equation}\label{PowerLaw}
\mathcal P_{\mathcal R}(k) = \frac{1}{8\pi^2\epsilon_*}\frac{H_{\text{inf}}^2}{M_P^2}\bigg(\frac{k}{k_*}\bigg)^{n_s(k)-1},
\end{equation}
where $k_*$ is the pivot scale, then the perturbation energy fraction is given by
\begin{equation}
\Omega_{\delta}(k) = \frac{4}{\pi}\int_k^{k_{\text{kin}}}d\ln \tilde k \frac{\tilde k}{k}\mathcal P_{\mathcal R}(k_*)\bigg(\frac{\tilde k}{k_*}\bigg)^{n_s(k)-1} \approx \frac{4}{\pi n_s(k_{\text{kin}})}\mathcal P_{\mathcal R}(k_{\text{kin}})e^{2\Delta N},
\end{equation}
where $k_{\text{kin}}/k = e^{2\Delta N}$ during kination. We see that the modes that dominate this energy fraction are the ones that first enter the horizon after inflation since they have the most time to be grow. $\Omega_\delta(k)$ reaches unity $11$ e-folds after the end of inflation \cite{Eroncel:2025bcb}, following which the cosmology would transition into the perturbation domination era, to be followed by the self-tracker.

During kination, the background scalar field evolves as 
\begin{equation}\label{kination}
\phi = \phi_0 + \sqrt 6 M_P \Delta N,
\end{equation}
while the exponential potential is given by
\begin{equation}
V(\phi) = V_0 e^{-\lambda \phi/M_P} = V_{\text{kin}} e^{-\sqrt 6 \lambda \Delta N},
\end{equation}
where $V_{\text{kin}}$ is the value of the potential at the start of the kination epoch. If the universe immediately transitions from inflation to kination, then the number of e-folds between the end of inflation and the potential reaching a particular IR-scale $V_{\text{IR}}$ is given by
\begin{equation}
\Delta N = \frac{1}{\lambda \sqrt 6} \ln \bigg(\frac{V_{\text{kin}}}{V_{\text{IR}}}\bigg).
\end{equation}
Kination occurs on exponential potentials as long as $\lambda > \sqrt 6$ \cite{Copeland:1997et} in which case for $V_{\text{IR}} = 10^{-20}V_{\text{kin}}$, we find that $\Delta N \leq 7.7$. In the case of the Large Volume Scenario \cite{Balasubramanian:2005zx} where $\lambda = \sqrt{27/2}$, the duration before reaching the minimum is $\Delta N = 5.1$. Therefore, if the scalar field perturbations are sourced entirely from slow-roll inflation, then the resulting perturbation amplitude is insufficient for the post-inflationary scenario on exponential potentials to reach the self-tracker before the scalar field reaches the minimum.

This does not mean that the self-tracker cannot be reached. In various scenarios, inflationary perturbations are enhanced near the end of inflation -- one popular examples is in the context of models of PBH formation. The existence of such enhanced perturbations would drastically cut down the time needed for the perturbations to catch up with the kination energy. In that case the self-tracker can be reached before reaching the minimum of the potential, which would solve the overshoot problem.

A source of radiation that could compete with the scalar field perturbations during kination is the radiation sourced by the de Sitter thermal bath during inflation, with a temperature of \cite{Gibbons:1977mu}
\begin{aleq}
T_{\text{dS}} = \frac{H_{\text{inf}}}{2\pi}.
\end{aleq}
If the perturbation contribution is subdominant compared to the thermal radiation energy density, then the tracker that would be reached would be a regular cosmological tracker as opposed to the self-tracker. Since inflation dilutes any particles gravitationally produced during inflation, the only such particles to survive would be those created near the end of inflation when the Hubble constant is approximately that found at the start of kination $H_{\text{kin}}$. The resulting thermal energy fraction at the start of kination, where $\bar \rho = 3H^2_{\text{kin}}M_P^2$, is given by
\begin{equation}
\Omega_\gamma = \frac{\rho_\gamma}{\bar \rho} = \frac{g_{\text{eff}}}{1440\pi^2} \frac{H^2_{\text{kin}}}{M_P^2},
\end{equation}
where $g_{\text{eff}}$ is the effective number of degrees of freedom gravitationally produced during inflation. $H_{\text{inf}}$ and $H_{\text{kin}}$ refer, respectively, to the Hubble scales at the times sixty e-folds before the end of inflation and at the start of kination: in slow-roll inflation these will be similar, but not identical. For the power law spectrum \eqref{PowerLaw} they are related by
\begin{equation}
H_{\text{kin}}^2 = H_{\text{inf}}^2 \bigg(\frac{k_{\text{kin}}}{k_*}\bigg)^{n_s(k_{\text{kin}})-1}.
\end{equation}

As both $\Omega_\gamma$ and $\Omega_\delta$ scale as radiation, whichever one started larger comes to dominate first. This primarily depends on the value of $g_{\text{eff}}$. In particular, the two components are equal when
\begin{equation}
g_{\text{eff}} = \frac{11520}{\pi}\frac{1}{rn_s(k_{\text{kin}})},
\end{equation}
where we set $\epsilon_* = r/16$. For $r = 0.01$ and $n_s(k_{\text{kin}}) \approx 1$, we find that $g_{\text{eff}} \approx 3.7\times 10^5$. Most particles in the Standard Model are not gravitationally produced since they are effectively massless compared to the inflationary Hubble scale where they couple conformally to gravity \cite{Parker:1969au}, hence $g_{\text{eff}} \sim \mathcal O(1)$. In extensions to the Standard Model, such as in string theory, one can get a large number of minimally coupled scalar fields that are produced giving $g_{\text{eff}} \sim \mathcal O(100) - \mathcal O(1000)$. However, even here the number of degrees of freedom would be insufficient to compete with the energy density in the perturbations, especially when taking into account the fact that the perturbations would need to be further amplified to reach the tracker before the scalar field reaches the minimum.

\subsection{Conclusion}

Tracker solutions are an interesting modification of the standard radiation-dominated cosmology. They are also important for avoiding the overshoot problem in string compactifications. By themselves, they are generally not viable late-time cosmologies as they correspond to rapidly evolving dark energy and can also lead to an unobserved fifth force. However, in the early universe they can guide the rolling modulus into its final vacuum while avoiding overshoot and runaway to infinity.

When tracker solutions based on exponential potentials are discussed in the literature, the background fluid is normally viewed as something entirely separate to the rolling scalar and its potential. A far more economical scenario would be one in which both come from the same source.

In this paper we have shown that such a scenario can indeed be realized. Through both analytical computation and numerical simulation using \CL, we have found that the self-perturbations of the rolling scalar field can themselves serve as a `background' radiation fluid. The numerical computations, which involve a full 3d simulation of the field, show that the results can be interpreted as a `background' averaged value of the field evolving in the presence of a `radiation fluid' sourced by the self-perturbations of the scalar field. The resulting fixed point of this solution has the energy proportions associated to conventional radiation trackers.

 If normalised to the inflationary power spectrum, it turns out that the initial amplitude of the self-perturbations would be too small to allow the system to reach the self-tracker solution before the rolling modulus reaches the minimum of the potential. However, if the inflationary perturbations are enhanced prior to the end of inflation (as happens for models of PBH formation), this problem can easily be avoided.

One area of future work would be to study the dynamics of the scalar self-tracker as the field reaches the minimum of the potential. As it does, the self-perturbations will convert from radiation perturbations to matter perturbations. The era of moduli domination can be long, potentially resulting in growth and collapse into bound structures (e.g. see \cite{Eggemeier:2021smj} for a recent numerical analysis). The spectrum of the self-perturbations may then imprint itself in the nature and spectrum of these structures.

\acknowledgments
We thank Fien Apers, Ethan Carragher, Cheng Cheng, Panagiotis Giannadakis, Lucien Heurtier, Eugene Lim and Noelia Sanchez-Gonzalez for discussions related to this topic. JC acknowledges support from the STFC consolidated grants ST/T000864/1 and ST/X000761/1 and is also a member of the COST Action COSMIC WISPers CA21106, supported by COST (European Cooperation in Science and Technology). MM is supported by the St John’s College Graduate Scholarship in partnership with STFC. EJC acknowledges support from the STFC Consolidated Grant ST/X000672/1. For the purpose of Open Access, the authors have applied a CC BY public copyright licence to any Author Accepted Manuscript version arising from this submission. 

\bibliography{biblist}

\end{document}